\documentstyle[prl,aps,epsf,multicol]{revtex}
\begin{document}

%%%%%%%%%%%FIGURES%%%%%%%%%%%%%%%%%%
%
\newcommand{\fig}[2]{\epsfxsize=#1\epsfbox{#2}}
% 
%
%%%%%%%%%DEUX COLONNES%%%%%%%%%%%%%%
%   
\newcommand{\passage}{%%
\end{multicols}\widetext\noindent\rule{8.8cm}{.1mm}%
  \rule{.1mm}{.4cm}} 
 \newcommand{\retour}{%%
 %        \hspace{.2cm}
\noindent\rule{9.1cm}{0mm}\rule{.1mm}{.4cm}\rule[.4cm]{8.8cm}{.1mm}%
         \begin{multicols}{2} }
 \newcommand{\unecol}{\end{multicols}}
 \newcommand{\deuxcol}{\begin{multicols}{2}}
%
%
%%%%%%%%%%%%%%%%%%%%%%%%%%%%%%%%%%%%
%
%
\newcommand{\beq}{\begin{equation}}
\newcommand{\eeq}{\end{equation}}
\newcommand{\beqa}{\begin{eqnarray}}
\newcommand{\eeqa}{\end{eqnarray}}
%
%\cst {\rm Cst}

%\setcounter{page}{1}

\tolerance 2000

\author{Baruch Horovitz{$^1$} and Pierre Le Doussal{$^2$} } 
\address{{$^1$} Department of Physics, Ben Gurion university, Beer Sheva
84105 Israel}
\address{{$^2$}CNRS-Laboratoire de Physique Th{\'e}orique de
l'Ecole Normale Sup{\'e}rieure,
24 rue Lhomond,75231 Cedex 05, Paris France.}

\title{Disorder Induced Transitions in Layered Coulomb Gases 
and Superconductors}
% \date{\today}
\maketitle

\begin{abstract}
A 3D layered system of charges with logarithmic interaction parallel
to the layers and random dipoles is studied via
a novel variational method and an energy rationale which reproduce the 
known phase diagram for a single layer. Increasing interlayer coupling
leads to 
successive transitions in which charge rods correlated in $N>1$
neighboring 
layers are nucleated by weaker disorder. 
For layered superconductors in the limit of
only magnetic interlayer coupling, the method 
predicts and locates a disorder-induced
defect-unbinding transition in the flux lattice.
While $N=1$ charges dominate there, $N>1$ disorder induced 
defect rods are predicted for multi-layer superconductors.
\end{abstract}

%\pacs{to be added}

%\narrowtext

\deuxcol

Topological phase transitions induced by quenched disorder
are relevant for numerous physical systems. Such transitions
are likely to shape the phase diagram of type II superconductors.
It was proposed \cite{tgpldbragg} that the flux lattice (FL) remains
a topologically ordered Bragg glass at low field,
unstable to the proliferation of dislocations 
above a threshold disorder or field, providing one
scenario for the controversial "second peak" line 
\cite{Kes,speakexp}. Another scenario \cite{H2}
is based on a disorder-induced decoupling transition (DT)
responsible for a sharp drop in the FL tilt modulus.
Furthermore, for the {\it pure} system, it was shown recently
\cite{Dodgson} that in the absence of Josephson coupling,
point "pancake" vortices, i.e vacancies and interstitials in the FL,
are nucleated at a temperature $T_{def}$, distinct from
melting above some field. It is believed that this
pure system topological transition merges with the 
thermal DT \cite{Daemen,H1} once the Josephson coupling 
is finite, being two anisotropic limits of the
same transition \cite{H3} (at which superconducting order
is destroyed while FL positional correlations are
maintained). Thus an interesting possibility is that a similar, 
but now disorder-induced, vacancy-interstitial unbinding
transition can be demonstrated in 3D layered 
superconductors, relevant to many layered 
and multilayer materials \cite{Kes,Bruynseraede}.

In 2D recent progress was made to describe
disorder induced topological transitions,
in terms of Coulomb gases of charges with
logarithmic long range interactions. It was shown
\cite{nattermann95,scheidl97,tang96,dcpld} that quenched 
random dipoles lead to a transition, via defect proliferation,
at a finite threshold disorder, even at $T=0$.

In this Letter we develop a theory for a
3D defect-unbinding transition in presence of disorder.
It is achieved for systems which can be mapped onto a
layered Coulomb gas with quenched random dipoles, in which the
interaction energy between two charges on layers $n$ and $n'$ is 
$2 J_{n-n'}\ln r$ with $r$ the charge separation parallel to the layers.
One physical realization is the FL
in layered superconductors \cite{Kes,Bruynseraede,Blatter}
with only magnetic coupling, for which we predict 
and locate the vacancy-interstitial unbinding
transition. Indeed, as we argue, disorder induced deformations
of the lattice result in random dipoles as seen by the defects.
To study this problem we  develop an efficient variational
method which allows for fugacity distributions,
known \cite{dcpld} to be important in 2D as they become
broad at low $T$. We test the method on a
single layer and reproduce the phase diagram, known from
renormalization group (RG) with a $T=0$ disorder threshold
$\sigma_{cr}=1/8$
\cite{footnote2}. For the 2-layer system we find that
above a critical anisotropy $\eta \equiv -J_1/J_0 = \eta_c = 1-
\frac{1}{\sqrt{2}}$
the single layer type transition is preempted by a transition induced
by bound states of two pancake vortices on the two layers with
$\sigma_{cr}<1/8.$ We develop a $T=0$ energy rationale by an
approximate mapping to a Cayley tree problem and find that it
reproduces the 2-layer result. Extension to many
layers with only nearest layer coupling shows a
cascade of transitions in which the number of correlated
charges on $N$ neighboring layers increases, while the critical
disorder decreases with $\eta$, with $N\rightarrow \infty$,
$\sigma_{cr}\rightarrow 0$ as $\eta \rightarrow 1/2$. 
Finally we consider arbitrary range $n _0$ for $J_n$ with the constraint
$\sum_n J_n =0$, as appropriate for layered superconductors.
For $N > n _0^2$ states with $\sigma_{cr} \sim n_0^2/ N \rightarrow 0$
are possible but only at exponentially large length scales for
$n _0 \gg 1$. Thus for layered
superconductors we expect that the N=1 state dominates and find
its phase diagram. Varying the system parameters by forming
multilayers reduces $n _0$ and allows for realization of the new
$N>1$ phases.

We study the Hamiltonian:

\begin{eqnarray}
&& {\cal H}= - \case{1}{2} \sum_{{\bf r} \neq {\bf r}'}\sum_{n,n'}2J_{n-n'}
 s_n({\bf r}) \ln ({\bf r}-{\bf r}') s_{n'}({\bf r}') \\
&& -
 \sum_{{\bf r},n} V_n({\bf r}) s_n({\bf r}) \label{H}
 \end{eqnarray}
 where $s_n({\bf r})=\pm 1, 0$ define the positions ${\bf r}$ of charges
on the $n$-th layer, $V_n({\bf r})$ is a disorder potential with long
range correlations $\overline{V_n({\bf q}) V_{n'}({-\bf q})}
= 4 \pi \sigma J_0^2 \Delta_{n-n'}/q^2$ with $\Delta_{0}=1$
(the short distance cutoff being set to unity).
For simplicity we start with uncorrelated disorder from layer to layer
$\Delta_{n-n'} = \delta_{nn'}$ with
\begin{eqnarray}
 \overline{[V_n({\bf r})-V_{n}({\bf r}')]^2}= 4 \sigma J_0^2
 \ln|{\bf r}-{\bf r}'| \label{corr}
\end{eqnarray}
representing quenched dipoles on each layer.
At $T=0$ the problem amounts to find minimal energy configurations 
of charges in a logarithmically correlated random potential. For a
single layer it was studied either using \cite{nattermann95,rem}
a ``random energy model'' (REM)
approximation \cite{footnote1}, or more accurately using a
representation in terms of directed polymers on a Cayley tree (DPCT)
\cite{tang96} shown to emerge \cite{dcpld} (as a continuum
branching process) from the Coulomb gas RG 
of the single layer problem. Schematically, the tree has independent 
random potentials (Fig. 1)
$v_i$ on each bond with variance $\overline{ v_i^2}=2 \sigma J_0^2$.
After $l$ generations one has $\sim e^{2 l}$ sites which are mapped
onto a 2D layer, i.e. two points separated by $r \sim e^{l}$ have
a common ancestor at the previous $l \approx \ln r$ generation.
Each point ${\bf r}$ has a unique path on the tree (DP) with
$v_1,...,v_l$ potentials and is assigned a potential $V({\bf
r})=v_1+...+v_l$. Since all bonds previous to the common ancestor
are identical $\overline{[V({\bf r})-V({\bf r}')]^2}=
2 \sum_{i=1}^{l}\overline{v_i^2}$
reproducing (\ref{corr}) on each layer. Exact
solution of the DPCT \cite{ct} yields the energy gained
from disorder $V_{min}=min_{\bf r} V({\bf r}) 
\approx - \sqrt{8 \sigma} J_0 \ln L$ for a volume $L^2$,
with only $O(1)$ fluctuations \cite{dcpld}, i.e
$- \sqrt{8 \sigma} J_0$ per generation $l=\ln L$.

\begin{figure}[htb]
\centerline{ \fig{6cm}{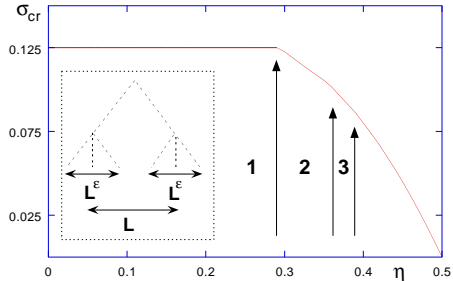} }
\caption{{\narrowtext  Critical disorder values with only
nearest neighbor coupling $J_1$ vs. the anisotropy $\eta=-J_1/J_0$.
Transitions between different $N$ phases are marked with arrows.
Inset: the Cayley tree representation (for $N=3$ neighboring layers)
with $+$ charges (at the tree endpoints) separated by $L^\epsilon$
along the layers, and separated by $L$ from the $N=3$ $-$ charges.}}
\label{fig1}
\end{figure}

Optimal energy configurations for $M$ coupled layers
are constructed considering $N$ neighboring layers
with a $+,-$ pair on each layer and no charges on the other layers.
We can take $J_0>0$ and $J_{n \neq 0} \leq 0$ 
so that {\em equal}
charges on different layers attract. The DPCT representation
now involves, on a single tree, $N$ $+$ polymers (each seeing different 
disorder) and $N$ $-$ polymers (each seeing opposite disorder $-v_i$
to their $+$ partner). A plausible configuration is that
the $+$ charges bind
within a scale $L^{\epsilon}$
($0 \le \epsilon \le 1$), so do the $-$ charges, while the $+$ to
$-$ charge separations define the scale $L$. Its tree representation
(Fig. 1) has $2N$ branches with $\epsilon \ln L$ generations,
i.e. an optimal energy of $-2 N \sqrt{8\sigma}J_0\epsilon \ln L$.
On the scale between $L^{\epsilon}$ and $L$ the $+$ charges 
act as a single charge with a potential $\sum_{n=1}^N V_n({\bf r})$
(the $N$ polymers share the same branch) of variance $N\sigma$
hence the optimal energy is $-2\sqrt{8N\sigma}J_0 (1-\epsilon)\ln L$.
The total disorder energy is
\cite{footnote3}:
\begin{equation}
E_{dis} \approx -2J_0\sqrt{8\sigma}[\epsilon N
+(1-\epsilon)\sqrt{N}]\ln L \,.
\end{equation}

The competing interaction energy $E_{int}$ is the sum of the 
one for the $+-$ pairs, $[2J_0N+4 \sum_{n=1}^{N}J_n(N-n)]\ln L$
and for the $++/--$ pairs, $-4\sum_{n=1}^N J_n(N-n)\epsilon \ln L$.
The total energy
$E_{tot}=E_{dis}+E_{int}$ being linear in $\epsilon$, its
minimum is at either $\epsilon=1$ or $\epsilon=0$.
Since $\epsilon =1$ implies 
that the $+$ charges unbind, it is
sufficient to consider $\epsilon=0$ with all $N \geq 1$, i.e. a 
rod with $N$ correlated charges has energy
(with $\eta_n =-J_n/J_0$):
\begin{equation}
E_{tot} = 2J_0N[1-2\sum_{n=1}^N \eta_n(1-\frac{n}{N})
-\sqrt{\frac{8\sigma}{N}}]
\ln L \,. \label{Etot}
\end{equation}
Disorder induces the $N$ vortex state at the critical value:
\begin{equation}
\sigma_{cr}=\frac{N}{8}[1-2\sum_{n=1}^N\eta_n (1-\frac{n}{N})]^2
\,.
\end{equation}
(i.e. $E_{tot}=0$). Consider first only nearest neighbor coupling
$\eta_l=\eta_1 \delta_{l 1}$. Then $\sigma_{cr}$ is minimal 
at $N=1$ with $\sigma_{cr}=1/8$ if $\eta _1<1-1/\sqrt{2}$.
For larger anisotropies successive $N$ states form at 
$1/(1-2\eta _1) = 1 + \sqrt{N(N-1)} \sim N$
with diverging $N$ as $\eta _1 \to \frac{1}{2}$ (Fig 1)
\cite{footnote9}.

Consider now $J_n$ of range $n_0$ constrained by
$\sum_n J_n=0$ as for the superconductor, e.g.
$\eta_n=\eta_1 e^{-(n-1)/n_0}$ for which
$\sigma_{cr}=(1-e^{-N/n_0})/8N(1-e^{-1/n_0})$.
For $n_0 \gg 1$, each $\eta _{n\ne 0}$ is small:
for $N \lesssim n_0$ the lowest $\sigma_{cr}$ is at $N=1$.
However, the combined strength of $N \approx n_0$ vortices being
significant $\sigma_{cr}$ has a maximum
and decreases back to zero for $N > n_0$ as $\sigma_{cr} \approx n_0^2/8N$. 
Hence $\sigma_{cr} \rightarrow 0$ as $N \rightarrow
\infty$ and any small disorder seems to nucleate such vortices.
This is because the perfect screening of the zero mode
$\sum_n J_n=0$ implies that an infinite 
charge rod has a vanishing $\ln r$ interaction; hence a
logarithmically correlated disorder is always dominant.

The realization of the large $N$ rods
depends, however, on the type of thermodynamic limit.
Adding to (\ref{Etot}) the core energy $E_c N$ and
minimizing yields a $N$-vortex scale
\begin{equation}
L\approx \exp
\{E_c\sqrt{N}/[2J_0(\sqrt{8\sigma}-\sqrt{8\sigma_{cr}})]\}
\label{L}\,.
\end{equation}
Hence as $\sigma\rightarrow 0$ such states are only
achievable when $L/N$ diverges
exponentially. Using $\sigma_{cr}\approx n_0^2/8N$,
for $N> n_0^2/8\sigma$ the
lowest scale $L$ in this range is achieved at $N=n_0^2/2\sigma$
and leads to a lower bound
$L_{min}\approx \exp [E_c n_0/4J_0 \sigma]$ for observing
large $N$ states with a given $\sigma <\case{1}{8}$.
For layered superconductors $E_c/J_0 \gg 1$ \cite{footnote7} and $n_0 \gg 1$
and this large $N$ instability occurs at unattainable
scales, thus $N=1$ dominates. One needs $n_0\approx 2-3$,
as in multilayers,
to realize the $N>1$ states.

To substantiate these results we develop a variational method
for $M$ layers which allows for fugacity distributions,
an essential feature in the one-layer problem.
Disorder averaging (\ref{H}) in Fourier using replicas yields:
\begin{eqnarray}
&& \beta {\cal H}_r= \case{1}{2 d^2} \int_k \int_q
s_a({\bf q},k)(G_0)_{ab}({\bf q},k)s_b^*({\bf q},k) \label{Hr} \\
&& + \beta E_c \sum _{{\bf r},n} s_{na}^2({\bf r}) \nonumber
\end{eqnarray}
where $(G_0)_{ab}({\bf q},k)=(4\pi /q^2)[g(k) \delta_{a b}- 
\sigma J_0^2 \beta^2
\Delta(k)]$, $g(k)=\beta J(k) = 
\beta d \sum_n J_n \exp(ikdn)$, $d$ the interlayer
spacing \cite{footnote5} (for uncorrelated layers 
$\Delta(k) = d$), $a,b=1,...,m$ are replica indices
and $m\rightarrow 0$ is to be carefully taken.
In transforming to a sine-Gordon Hamiltonian \cite{H3}
it is crucial to keep {\it all} charge fugacities \cite{dcpld},
which yields:
\begin{eqnarray}
&& \beta {\cal H}_{SG} =\case{1}{2} \int_{k q}
\chi _a({\bf q},k)(G_0)^{-1}_{ab}\chi^*_b({\bf q},k) \nonumber \\
&& - \sum_{{\bf r}} \sum_{{\bf s} \neq {\bf 0}} Y[{\bf s}]
\exp( i {\bf s} \cdot {\mbox{\boldmath $\chi$}}({\bf r}) ) \quad .
\label{sg}
\end{eqnarray}
>From now on ${\bf s} = \{ s_{na} \}_{n=1,.M,a=1,.m}$
is an integer vector both in layer label and replica 
space (i.e. of length m M)
of entries $0, \pm 1$ and the summation is over all such non null
vectors (also $\chi({\bf r}) \equiv \{ \chi_{n,a}({\bf r}) \}$, 
${\bf s} \cdot {\mbox{\boldmath $\chi$}}
=\sum_{na} s_{na} \chi_{na}$).
We now look for the best gaussian approximation of (\ref{sg})
with propagator $G^{-1}_{ab}({\bf q},k)=(G_0)^{-1}_{ab}({\bf q},k) +
\sigma _c(k) \delta _{ab} + \sigma_0(k)$. The bare fugacity 
being $Y[{\bf s}]=\exp{(-\beta E_c \sum_{n,a} s_{n,a}^2)}$
the naive approach would be to restrict to charges ${\bf s}$ with a
single non zero entry, leading to a uniform fugacity term
$ - y \sum_{{\bf r},n,a} \cos({\mbox{\boldmath $\chi$}_{na}}({\bf r}))$
and a diagonal $k$-independent replica mass term. Instead we
keep {\it all} composite charges ${\bf s}$, which allow for
variational solutions with off diagonal and $k$-dependent replica mass terms.
This corresponds respectively to fluctuations of fugacity 
and $N >1$ charge rods being generated and becoming relevant
as also seen from RG.
The variational free energy is ${\cal F}_{var}={\cal F}
_0+\langle {\cal H}_{SG}-{\cal H}_0\rangle_0$ where
$\langle...\rangle$ is an average using $\beta {\cal H}
_0=\case{1}{2}\int_{{\bf q},k}\chi_a({\bf q},k)G_{ab}({\bf
q},k)\chi^*_b({\bf q},k)$ and 
$\beta {\cal F}_0=-\case{1}{2} \text{tr} \ln G$.
The Gaussian average 
$F[{\bf s}] \equiv Y[{\bf s}]
\langle \exp{i{\bf s} \cdot {\mbox{\boldmath
$\chi$}}({\bf r})}\rangle_0$ yields:
\begin{eqnarray}
&& F[{\bf s}] = \exp\{-\case{1}{2 d^2} 
\int_k \sum_{ab} (\tilde{G}_c(k) \delta_{ab} - A(k)) 
s_a(k) s^*_b(k)\}
\nonumber \\
&& G_c(k) =  g(k)\ln[\Lambda/(4\pi g(k)\sigma_c(k))] \\
&& A(k) = \sigma \beta^2 J_0^2 
\Delta(k) (G_c(k)/g(k) - 1) + g(k)\sigma_0(k)/\sigma_c(k) \nonumber
\end{eqnarray}
where $s_a(k)=d \sum_n s_{na} e^{i k n d}$, 
$\tilde{G}_c(k)=G_c(k)+2 \beta E_c d$, $\Lambda$ the UV cutoff on $q^2$.
$F_{var}$ is minimized by $\sigma_c(k)\delta_{ab}+\sigma_0(k)=
\Lambda d^{-2} \sum_{{\bf s}} s_{a}(k) s_{b}^*(k) F[{\bf s}]$.
Writing the $A(k)$ term as an average over
$M$ random gaussian fugacities ${w_{k}}$:
\begin{equation}
\exp{\{\case{1}{2}|\sum_{a} s_a(k)|^2 A(k)\}}=
\langle \exp{w_{k}\sum_{a} s_a(k)} \rangle_{w}
\end{equation}
where $\langle...\rangle_{w}=\prod_k \int...
e^{-|w_{k}|^2/2 A(k)} d^2 w_{k}/\sqrt{2\pi A(k)}$,
allows to perform the exact sum on replicas yielding
$\sum_{{\bf s}} F[{\bf s}] = \langle Z^m\rangle_{w}$
with $Z=\sum_{\{s_n=0,\pm 1\}}
\exp(- \frac{1}{2 d^2} \int_k \tilde{G}_c(k) |s(k)|^2 
+ \frac{1}{d} \int_k w_{k} s^*(k))$.
The variational equations for $m\rightarrow 0$ become
\cite{us_inprep}
\begin{equation}
\sigma_c(k)= \Lambda \langle \frac{\partial^2 \ln Z}{\partial w_{k}
\partial w^*_{k}} \rangle_{w} \,; {\mbox {\hspace{2mm}}}
\sigma_0(k)= \Lambda \langle |\frac{\partial \ln Z}{\partial 
w_{k} }|^2 \rangle_{w} \label{consistency}
\end{equation}
For a single layer $k=0$ and $Z=1+e^{u+w}+e^{u-w}$,
$2 u=-\tilde{G}_c(0)/d$, $w=w_0$ is a trinomial.
(\ref{consistency}) can be solved for the critical line
where $\sigma_c(0) \rightarrow 0$.
The phase diagram shown in Fig. 2 (full line) reproduces
precisely recent RG results. The variational scheme, allowing for
all replica charges ${\bf s}$, therefore treats disorder
correctly. For two layers $k d=0,\pi$
we need two fugacity distributions
$w_0,w_\pi$ and $Z$ is a "ninomial",
i.e. $Z=1+$ eight exponentials involving
$G_c(0)$, $G_c(\pi)$. Focusing on the low $T$
boundary, where $\sigma_c(\pi)\sim
[\sigma_c(0)]^{\alpha}\rightarrow 0$ we find \cite{us_inprep} either (i) $\alpha
=1$ for $\eta_1 < \eta_c = 1-1/\sqrt{2}$, representing decoupled layers, or
(ii) $\alpha \rightarrow \infty$ for $\eta_1
>\eta_c$, representing a $++$ bound states on the two
layers. 
The $T=0$ energy rationale is therefore reproduced. The phase
diagram for two layers with $\eta_c<\eta<1/2$ is shown
in Fig. 2 \cite{footnote2}

\begin{figure}[htb]
\centerline{ \fig{6cm}{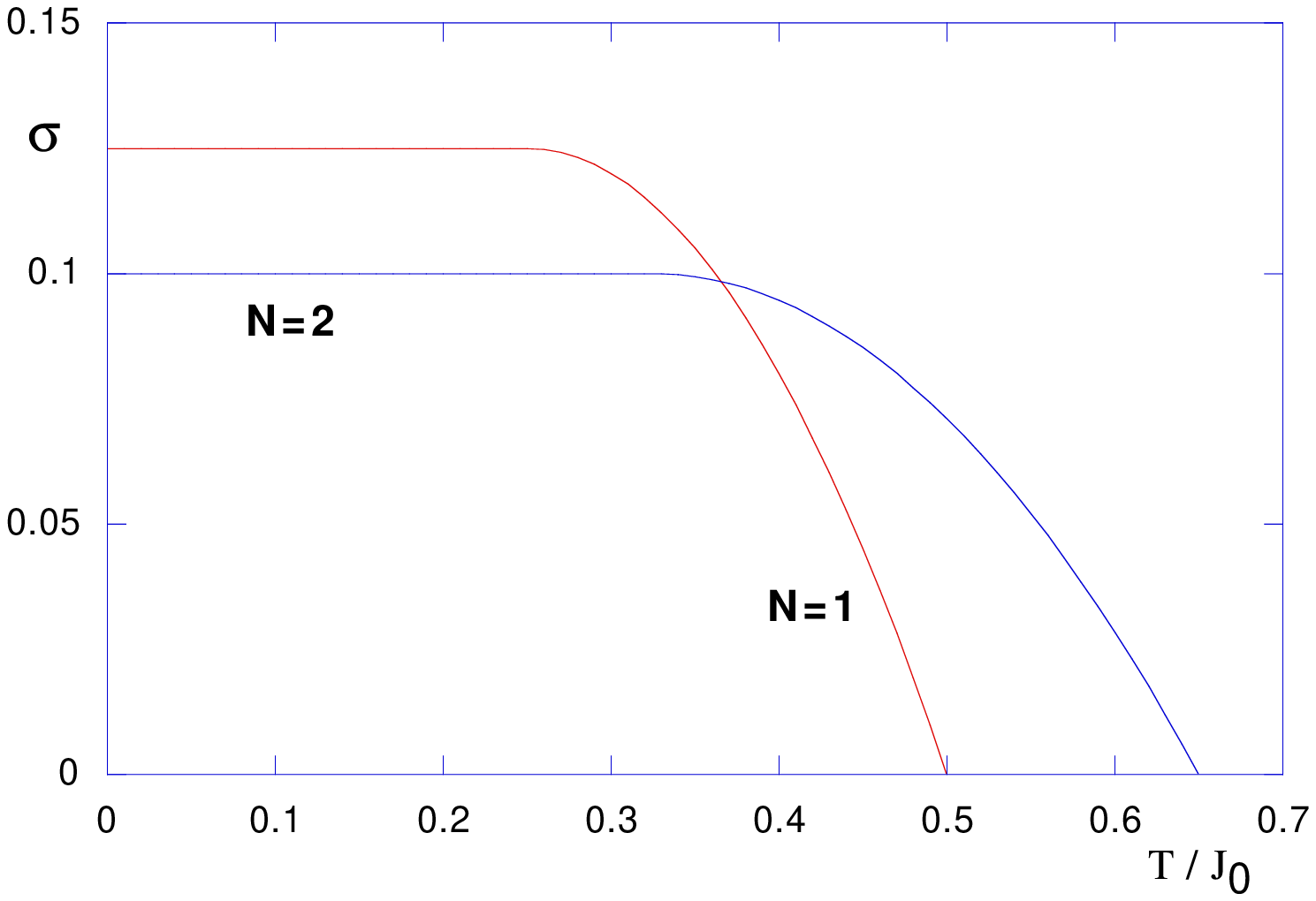} }
\caption{{\narrowtext  Phase diagram for the onset of 
the $N=1,2$ instabilities for anisotropy $\eta=0.35$. At low $T$
two distinct transitions are possible, the first being
to the rod $N=2$ phase. At high $T$
the independent layer $N=1$ transition dominates}}
\label{fig1}
\end{figure}

For any number of layers one obtains a simple $N$ rod solution
by restricting the sum over ${\bf s}$ in (\ref{sg}) to
a subclass of charges of the form 
$s_{na}= s_a \sum_{j=1,N} \delta_{n,n'+j-1}$. The variational
solution, of the form $\sigma_c(k)=\sigma_c \phi_N(k)$, reduces
to an effective one layer problem, in term of the
structure factor of the rod 
$s_a(k) s_b^*(k) = \phi_N(k) \equiv \sin^2(N k d/2)/\sin^2(k d/2)$.
The $N$ rod becomes critical at:
\begin{eqnarray}
\sigma_{cr} = (\int_k \phi_N(k) J(k))^2/(8 J_0^2 \int_k \phi_N(k) \Delta(k))
\label{critical}
\end{eqnarray}
a formula which can equivalently be obtained within the
Cayley tree rationale. Indeed for any correlations $\Delta_n$,
the energy of the $\epsilon =0$ configurations is still
given by (\ref{Etot}) replacing 
$\sigma \to \sigma (1 + 2 \sum_{n=1}^N \Delta_n (1-n/N))$).
(\ref{critical}) reproduces both single (using $\phi(0)=N$) and two layer
results. Finally, for
many layers and weak interlayer coupling (e.g. 
$\eta_1<1-1/\sqrt 2$ in Fig.1) the $N=1$ transition dominates
and occurs at $\sigma_{cr}=1/8$ ((\ref{critical}) using $\phi_1(k)=1$).

As a direct application we consider a flux lattice in
a layered superconductor with no Josephson coupling and a magnetic
field $B$ perpendicular to the layers. 
The FL is composed of pancake vortices
displaced from the
$p$-th line position ${\bf R}_p$ at the $n$-th layer into 
${\bf R}_p+{\bf u}_p^n$. The
defects $s_n({\bf r})$ couple to the lattice via ${\cal
H}_{vac}=\sum_{{\bf r},p,n,n'} s_{n'}({\bf r})G_v({\bf R}_p+{\bf
u}_p^n -{\bf r}, n'-n)$ where, in Fourier 
\cite{H3} $G_v({\bf q},k)=(\phi _0^2d^2/4\pi
\lambda_{ab}^2q^2)/[1+f({\bf q},k)]$ where $f({\bf
q},k)=(d/4\lambda_{ab}^2q)\sinh qd /[\sinh ^2(qd/2)+ \sin
^2(kd/2)] $; $\phi_0=B a^2$ is the flux quantum,
$a$ the FL spacing, $\lambda_{ab}$ the
penetration length along the layers.
 To 0-th order in ${\bf u}_p^n$ the defects feel
a periodic potential fixing their position in a unit
cell, hence $s({\bf q},k)$ involve only $|q|<1/a$.

In the limit $q \to 0$ the longitudinal modes,
to which defects couple,
have for (tilt) elastic energy \cite{Goldin} 
${\cal H}_{el}=\case{1}{2 d^2 a^4} \int_{kq} D(k)|{\bf u}_L({\bf q},k)|^2$
with $D(k)=\case{1}{2}\sum_{{\bf Q}\neq 0
}[G_v({\bf Q},k)-G_v({\bf Q},0)]+G_v(k)$ where ${\bf Q}$ are
reciprocal wavevectors of the lattice and $G_v(k)=\lim
_{q\rightarrow 0}G_v({\bf q},k)q^2=\phi_0^2d^2k_z^2/[4\pi
(1+\lambda_{ab}^2k_z^2)]$ and $k_z=(2/d)\sin (kd/2)$. The sum on
${\bf Q}$ is due to the high momentum components of the magnetic
field and is responsible for the non-perfect screening of the
defect interaction and to a finite $T_{def}$. Minimizing ${\cal
H}_{vac}+{\cal H}_{el}$ yields 
${\bf u}_{vac}({\bf q},k)=i{\bf q}s({\bf q},k)G_v(k)a^2/D(k)q^2$
and (\ref{Hr}) with:
\begin{equation}
g(k)=\beta G_v(k)[1-G_v(k)/D(k)]/4\pi \label{g}
\end{equation}
Thus the long range interaction is $\sim \ln r$ and its
coefficient determines $T_{def}=2 J_0$ (via $\int_k g(k)=2$).
Since $\int_k G_v(k) \sim \phi_0^2 d/\lambda_{ab}^2$,
the scale of the melting transition
\cite{Blatter}, the defect transition occurs before melting and
can thus be consistently described only if $D(k)-G_v(k)\ll D(k)$.
This is possible if
either $d\ll a \ll \lambda_{ab}$ where 
$g(k)=\beta d\tau' \ln (1+a^2k_z^2/4\pi)$ with 
$\tau'=\phi_0^2da^2/(128\pi^3 \lambda_{ab}^4)$ and \cite{Dodgson}
$T_{def}=\tau' ln (a/d)$ or for $d >
a$ where $g(k)=\beta (d^4/a)\tau' k_z^2e^{-2\pi
d/a}$ leading to $T_{def}=4(d/a) \tau'
e^{-2\pi d/a}$. Remarkably $D(k)-G_v(k)\ll D(k)$ also 
yields that the long range response ${\bf u}_{vac}({\bf
r})\sim a^2 {\bf r}/r^2$ to a vacancy 
at ${\bf r}=0$ is confined to the same layer.

Point disorder deforms the flux lattice, producing quenched
dipoles coupling to our defects. Expansion of the disorder 
energy, valid below the Larkin length 
\cite{tgpldbragg}, and minimization together with ${\cal
H}_{vac}+{\cal H}_{el}$ yields readily (\ref{H}). A more
general argument, valid at all scales, treats
$u_{vac}$ as a small perturbation around the Bragg glass
configuration. Systematic expansion of the free energy
$F = F_{BG} + \frac{1}{d^2 a^2} \int_{q k} i q s(q,k) G_v(q,k) 
\langle u(q,k) \rangle_{s=0}
+ O(s^2)$ in defect density in a given disorder configuration
shows that a defect feels a logarithmically correlated
random potential $V_l({\bf r})$ as in (\ref{H}, \ref{Hr})
with $\sigma J_0^2 \Delta (k) = 
G_v(k)^2 \lim_{q \to 0} C_{BG}(q,k)/4 \pi d^2 a^4$
where $C_{BG}({\bf r},l)=\overline{\langle u^L_0({\bf 0}) \rangle 
\langle u^L_l({\bf r}) \rangle}$
is the correlation in the unperturbed
Bragg glass
$C_{BG}(0,k) \sim 1/(c^2_{44} (k^4 + R_c^{-1} k^3)$,
$R_c$ a Larkin length along $c$ \cite{tgpldbragg}. It yields a $k$-independent $\Delta (k)$
for $k>1/R_c$ while $\Delta (k) \sim k$ for $k < 1/R_c$.

Applications to FL depends on the interlayer form of (\ref{g})
of range $n_0\approx a/d$ for large $a/d$. Remarkably $g(k=0)=0$,
i.e perfect screening holds as in 2D \cite{Dodgson}.
Hence $\sum _n J_n=0$ and as $n_0$ is reduced $J_0, J_1$ dominate the sum, i.e.
$\eta_1 \rightarrow \case{1}{2}$ when $d \gg a$. One finds that
$\eta_1$ crosses the critical value $1-1/\sqrt{2}$ 
when $d/a\approx 1$, depending weakly on 
$a/\lambda_{ab}$. We thus propose that FL in
multilayer superconductors, where $d>a$ can be achieved, can show
a rich phase diagram with $N>1$ phases. 
In layered superconductors $a/d\approx 10-100$ \cite{Kes}
and the $N=1$ transition at $\sigma_{cr}=1/8$
dominates for realistic sizes. The disorder-induced
decoupling transition, neglecting defects, predicted \cite{H1}
at $\sigma_{dec}=2$ is thus above the defect transition
(with $B \sim \sigma$) in the $B-T$ plane (similarly thermal 
decoupling occurs at $T_{dec}= 8 T_{def}$ for $d\ll a\ll
\lambda$). A natural scenario is again of a single
transition at $\sigma_c$ varying from
$2$ to $1/8$ as the bare Josephson coupling is
reduced, e.g. by increasing $d$ in multilayers.

In conclusion, we developed a variational method and a Cayley
tree rationale to study layered Coulomb gas. The results are relevant
to flux lattices where we find the phase boundaries and
propose new $N>1$ phases for $d\gtrsim a$. The present methods
may be useful for other 2D disordered systems, such as quantum Hall.

This work was supported by the French-Israeli program Arc-en-ciel
and by the Israel Science Foundation.

%\vspace{20cm}
\unecol
%\end{references}


\begin{thebibliography}{999}

%\begin{references}

\bibitem{tgpldbragg}
T. Giamarchi, P. Le Doussal, Phys. Rev. B {\bf 52} 1242 (1995)
and Phys. Rev. B {\bf 55} 6577 (1997).
\bibitem{Kes} P. H. Kes, J. Phys. I
France {\bf 6}
2327 (1996).
\bibitem{speakexp} B. Kaykovich et al. Phys. Rev. Lett {\bf 76}
2555 (1996), K. Deligiannis et al. Phys. Rev. Lett {\bf 79}
2121 (1997).
\bibitem{H2} B. Horovitz, cond-mat/9903167, Phys. Rev. B {\bf 60} R9939
(1999).
\bibitem{Dodgson} M. J. W. Dodgson, V. B. Geshkenbein and G.
Blatter Phys. Rev. Lett. {\bf 83} 5358 (1999). 
\bibitem{Daemen} L. Daemen et al., Phys. Rev. Lett. {\bf 70}, 1167
(1993)
\bibitem{H1} B. Horovitz and T. R. Goldin, Phys. Rev. Lett. {\bf
80},1734 (1998).
\bibitem{H3} B. Horovitz, Phys. Rev. B{\bf 47}, 5947 (1993)
\bibitem{Bruynseraede} Y. Bruynseraede et al., Phys. Scr. {\bf T42},
37 (1992).

\bibitem{nattermann95} T. Nattermann et al., J. Phys. I (France) {\bf 5}, 565 (1995)

\bibitem{scheidl97} S. Scheidl, Phys. Rev. {\bf B 55}, 457 (1997)

\bibitem{tang96} L. H. Tang, Phys. Rev. {\bf B 54}, 3350 (1996).

\bibitem{dcpld} D. Carpentier and P. Le Doussal, 
Phys. Rev. Lett. {\bf 81} 2558 (1998), cond-mat/9908335
and in preparation.

\bibitem{Blatter} G. Blatter et al. Rev. Mod. Phys.
{\bf 66} 1125 (1994).

\bibitem{footnote2} these phase diagrams are exact
in terms of {\it renormalized} parameters $\sigma_R$, $g_R(k)$
as seen from RG studies\cite{dcpld,us_inprep}.

\bibitem{rem} B. Derrida Phys. Rev. B {\bf 24} 2613 (1981)

\bibitem{footnote1} i.e. replacing the $V({\bf r})$ by $L^2$ variables
{\it uncorrelated} in ${\bf r}$, with the same on-site 
variance $\overline{V^2({\bf r})} \sim 2 \sigma J_0^2 \ln L$
also yielding \cite{rem} $V_{min} \sim - \sqrt{8 \sigma} J_0 \ln L$.

\bibitem{ct} B. Derrida, H. Spohn, J. Stat. Phys. {\bf 51} 817 (1988).

\bibitem{footnote3} an upper bound which 
can be argued to be exact \cite{us_inprep}.

\bibitem{us_inprep} B. Horovitz and P. Le Doussal in preparation.

\bibitem{footnote9} if $\eta _1>\case{1}{2}$ the defects form a lattice
even without disorder.

\bibitem{footnote7} $E_c/J_0 \approx (\lambda_{ab}/a)^2 /\ln(a/d)$ in that case.

\bibitem{Goldin} T. R. Goldin and B. Horovitz, Phys. Rev. B{\bf
58}, 9524 (1998).

\bibitem{footnote5} with $\int_q \equiv \int
\frac{d^2q}{(2\pi)^2}$, $\int_k \equiv \frac{1}{M d}
\sum_k \to \int_{-\pi/d}^{\pi/d} \frac{dk}{2 \pi}$ for
large $M$, $s_a({\bf q},k)=d \sum_{n,{\bf r}} 
s_{na}({\bf r}) e^{i {\bf q} {\bf r} + i k d n}$ and
$\beta=1/T$.

\end{thebibliography}
\end{document}